\documentclass[preprint,12pt]{elsarticle}
\journal{Computer Physics Communications}
\usepackage{a4wide,epsfig,psfrag,amsmath,amssymb,scalefnt}
\usepackage[dvipsnames]{xcolor}
\usepackage{amsmath,comment,braket}
\usepackage{placeins}
\usepackage{breqn}
\usepackage{slashed}
\usepackage{comment}
\usepackage{subcaption}
\usepackage{hyperref}

\usepackage{listings}
\graphicspath{ {figs/} }

\newcommand{\dd}{\mathrm{d}}

\allowdisplaybreaks

\begin{document}

\begin{frontmatter}

\title{\mbox{}\hfill{\footnotesize LTH 1406, P3H-25-035, TTP25-017, ZU-TH 40/25}
\\[.5em]
{\tt ggxy}: a flexible library to compute gluon-induced cross sections}

\author[a]{Joshua Davies}
\ead{J.O.Davies@liverpool.ac.uk}

\author[b]{Kay Sch\"onwald}
\ead{kay.schoenwald@physik.uzh.ch}

\author[c]{Matthias Steinhauser}
\ead{matthias.steinhauser@kit.edu}

\author[c]{Daniel Stremmer}
\ead{daniel.stremmer@kit.edu}


\address[a]{
  Department of Mathematical Sciences, University of
  Liverpool, Liverpool, L69 3BX, UK
  }
\address[b]{
  Physik-Institut, Universit\"at Z\"urich, Winterthurerstrasse 190,
  8057 Z\"urich, Switzerland
}
\address[c]{
  Institut f{\"u}r Theoretische Teilchenphysik,
  Karlsruhe Institute of Technology (KIT),\\
  Wolfgang-Gaede Stra\ss{}e 1, 76131 Karlsruhe, Germany
}

\begin{abstract}

  We present the library {\tt ggxy}, written in {\tt C++}, which can be
  used to compute partonic and hadronic cross sections for gluon-induced
  processes with at least one closed heavy quark loop.  
  It is based on analytic ingredients which avoids, to a large extent,
  expensive numerical integration. This results in significantly shorter
  run-times than other similar tools. 
  Modifying input parameters, changing the renormalization scheme
  and varying renormalization and factorization scales is straightforward.

  In Version~1 of {\tt ggxy} we implement all routines which are needed to
  compute partonic and hadronic cross sections for Higgs boson pair production
  up to next-to-leading order in QCD. We provide flexible interfaces
  and allow the user to interact with the built-in amplitudes at
  various levels.


\vspace*{1em}

\noindent \textbf{PROGRAM SUMMARY}

\begin{small}
\flushleft
\noindent
{\em Program title:} {\tt ggxy}\\
{\em Developer's repository link:} \url{https://gitlab.com/ggxy/ggxy-release}\\
{\em Licensing provisions:} GNU General Public License Version 3\\
{\em Programming language:} {\rm C++} and {\rm Fortran}\\
{\em External routines/libraries used:} {\tt avhlib}, {\tt boost}, {\tt Collier}, {\tt CuTtools}, {\tt eigen}, {\tt LHAPDF}, {\tt lievaluate}, {\tt OneLOop}, {\tt Recola}, {\tt CRunDec}\\
{\em Nature of problem:} The computation of partonic and hadronic cross sections for 
gluon-induced processes. In Version~1, the Higgs boson pair production process
is implemented at next-to-leading order in Quantum Chromodynamics.\\
{\em Solution method:} For the virtual corrections, deep expansions around the
forward and high energy limit are used.\\
{\em Restrictions:} The run-times depend crucially on the requested precision.
Results at the per-mille level can be obtained in about $30$ minutes. \\
{\em References and Links:} are provided in the paper 
\end{small}


\end{abstract}


\end{frontmatter}


\thispagestyle{empty}

\newpage

\vspace{1cm}

\tableofcontents

\newpage


\section{Introduction}

At the Large Hadron Collider (LHC) at CERN, gluon-induced processes have
a comparatively large cross section and are thus important for various
phenomenological analyses, in particular in the context of the Higgs boson
(see, e.g., Ref.~\cite{LHCHiggsCrossSectionWorkingGroup:2016ypw}).
This is also true for Higgs boson pair production, a process which already
receives a lot of attention from the experimental groups.
It will further
increase during the high-luminosity phase of the LHC,
since it is the only way to access the self-interaction of the 
Higgs boson directly.

Higgs boson pair production
is one of the first gluon-initiated loop-induced $2 \to 2$ processes which has been computed to
next-to-leading order (NLO) in QCD without relying on any 
approximation~\cite{Borowka:2016ehy,Borowka:2016ypz,Baglio:2018lrj}. This was an
important milestone since the NLO corrections amount to
almost 100\% of the LO and are thus very important. In addition, there is a large
theoretical uncertainty due to the dependence on the value and renormalization scheme
of the top quark mass.

The first calculation with exact $m_t$ dependence at NLO
from~\cite{Borowka:2016ehy,Borowka:2016ypz} suffered from large run-times.  This
has been circumvented by generating an interpolation grid for the virtual
amplitude~\cite{hhgrid}. However, this approach has the disadvantage that the top quark
and Higgs boson masses have fixed values. As a consequence, analyses using the
$\overline{\rm MS}$ definition of the top quark mass with an energy-dependent
renormalization scale are prohibitively expensive.  In~\cite{Baglio:2018lrj}
such an analysis has been performed, however no (detailed) timings are provided in
the paper.

A possibility to reduce or even eliminate the bottlenecks related to the
run-time and the lack of flexibility to modify numerical input values is to
incorporate analytic results from expansions in various phase-space regions.
In a first approach in this direction, the
analytic results obtained in the high-energy region were combined with
the purely numerical results of \cite{hhgrid}, which are then only needed in a restricted region
of phase space~\cite{Davies:2019dfy}.  In a later work the expansion around
small transverse Higgs boson momenta has been combined with the high-energy
expansion which completely avoids expensive numerical
calculations~\cite{Bellafronte:2022jmo,Davies:2023vmj}.

In this work we implement the results of~\cite{Davies:2023vmj} in a fast and
flexible {\tt C++} library: {\tt ggxy}. We supplement these results by our
implementation of the real radiation contribution and the possibility to
compute differential and total hadronic cross sections.  Higgs boson pair
production serves as a sample process. The structure of {\tt ggxy} is such
that it can easily be extended to other processes in the future.

The remainder of the paper is structured as follows: In Sections~\ref{sec::virt} and~\ref{sec::real} we comment
on the virtual and real corrections to $gg\to HH$, respectively, and
put special emphasis on the implementation in ${\tt ggxy}$.
Section~\ref{sec::ggxy} provides details on the
installation and the structure of {\tt ggxy}. We describe the dependencies and provide a manual for using the functions implemented in {\tt ggxy}.
Section~\ref{sec::examples} contains various examples which demonstrate the various ways to use {\tt ggxy}. The user is invited to adapt the
examples to their own purpose. We conclude in Section~\ref{sec::concl}.


\section{\label{sec::virt}Virtual corrections}

The virtual corrections to Higgs boson pair production have the same
kinematics as the Born process. They are conveniently parameterized in terms of
two form factors reflecting the two possible tensor structures for
$gg \to HH$. We follow the notation introduced in \cite{Davies:2023vmj} and
refer to this paper for details.  For convenience we summarize in the
following only those formulae which are relevant for the implementation in
{\tt ggxy}.

We introduce the perturbative expansion of 
the form factors
$F_1$ and $F_2$ as
\begin{eqnarray}
  F &=& F^{(0)} + \left(\frac{\alpha_s(\mu_r)}{\pi}\right) F^{(1)}
        + \left(\frac{\alpha_s(\mu_r)}{\pi}\right)^2 F^{(2)} 
        + \cdots
  \,,
  \label{eq::F}
\end{eqnarray}
and decompose them into ``triangle'' and ``box'' form factors
($k=0,1,\ldots$)
\begin{eqnarray}
  F_1^{(k)} &=& \frac{3 m_H^2}{s-m_H^2} F^{(k)}_{\rm tri}+F^{(k)}_{\rm box1}
                + F^{(k)}_{\rm dt1}\,, \nonumber\\
  F_2^{(k)} &=& F^{(k)}_{\rm box2} + F^{(k)}_{\rm dt2}\,.
                \label{eq::F_12}
\end{eqnarray}
$F^{(k)}_{\rm dt1}$ and $F^{(k)}_{\rm dt2}$ denote the contribution from
one-particle reducible double-triangle diagrams, see e.g.~Fig.~1(f) of
Ref.~\cite{Davies:2019dfy} (note that $F^{(0)}_{\rm dt1} = F^{(0)}_{\rm dt2} = 0$).
Analytic results for the leading-order form
factors are available from~\cite{Glover:1987nx,Plehn:1996wb}. The two-loop
triangle form factors have been computed in
Refs.~\cite{Harlander:2005rq,Anastasiou:2006hc,Aglietti:2006tp}, but we refrain from implementing them in {\tt ggxy} since the analytic expansions provide excellent approximations.
The exact results
for the double-triangle contribution can be found in~\cite{Degrassi:2016vss}, and are implemented in {\tt ggxy}.

In the pre-factor of $F_{\rm tri}^{(k)}$ it is possible to identify the trilinear self-coupling of the Higgs boson, $\lambda_3$, present in the Higgs potential
\begin{eqnarray}
   V(H) &=& \frac{m_H^2}{2}H^2 + \lambda_3 v H^3 + \frac{\lambda_4}{4}H^4\,,
\end{eqnarray}
where $v$ is the vacuum expectation value and $H$ is the Higgs boson field.
In the Standard Model we have
$\lambda_3^{\rm SM}=\lambda_4^{\rm SM}=m_H^2/(2v^2)$.
Deviations from this value are often parametrized
in the so-called $\kappa$ framework by introducing
$\kappa_\lambda=\lambda_3/\lambda_3^{\rm SM}$.
In {\tt ggxy} it is possible to choose a value for
$\kappa_\lambda\not=1$.
The current experimental bounds for 
$\kappa_\lambda$ from {\tt ATLAS} and {\tt CMS}
are
$-1.2<\kappa_\lambda<7.2$~\cite{ATLAS:2024ish}
and
$-1.39<\kappa_\lambda<7.02$~\cite{CMS:2024ymd}, respectively.

For the virtual corrections we introduce
the Mandelstam variables $s,t$ and $u$ which are defined in the usual
way with $s+t+u=2m_H^2$, and we also need the transverse momentum
of the Higgs bosons which is given by $p_T^2 = (tu-m_H^4)/s$.

For the virtual corrections we implement the (semi-)analytic expansions around
the forward and the high-energy limits as obtained in
Ref.~\cite{Davies:2023vmj}, where for the form factors $F_1$ and $F_2$
expansions in $t$ up to order $t^5$ and in $m_t$ up to order
$m_t^{120}$ have been computed.\footnote{Results to lower expansion
  depths in the high-energy expansion have been obtained in
  Refs.~\cite{Davies:2018ood,Davies:2018qvx}.  In the forward limit three
  expansion terms have been computed in Ref.~\cite{Bonciani:2018omm} in the
  context of the expansion in the transverse momentum of the Higgs boson,
  $p_T$.}  In both kinematic limits we expand in $m_H$ up to quartic order.
Although this is the limitation in precision of our approach, it is more than sufficient for all practical purposes. For higher values of the partonic centre-of-mass energy $\sqrt{s}$
and of $p_T$ the uncertainty is (far) below the percent level; close to the production
threshold $\sqrt{s}\approx 250$~GeV it can be larger
(in particular for $F_2$), however, there the
numerical values of the form factors are small.  Overall it has been
shown~\cite{Davies:2023vmj} that the expansions approximate the (exact)
numerical results with high precision.

In {\tt ggxy} ultraviolet renormalized and infrared (IR) subtracted form
factors are implemented. For the renormalization we apply the 
$\overline{\rm MS}$ scheme for $\alpha_s$ and renormalize the
external gluon field on-shell. For the top quark mass
there is the option to choose the on-shell ($M_t$) or 
$\overline{\rm MS}$ scheme ($\overline{m}_t(\mu_t)$), where
$\mu_t$ is the corresponding renormalization scale.
Our final results are expressed in term of $\alpha_s^{(5)}(\mu_r)$.

The subsequent subtraction of the remaining IR poles in $\epsilon$ leads
to finite form factors. Since there are different schemes concerning the
subtraction we provide explicit expressions for our implementation.
We obtain the finite form factors via
\begin{eqnarray}
  F^{(1),\rm fin} &=& F^{(1)} - \frac{1}{2} I^{(1)}_g F^{(0)}\,,
  \label{eq::FF_IR}
\end{eqnarray}
where the quantities on the right-hand side are ultraviolet-renormalized and
$I^{(1)}_g$ is given by~\cite{Catani:1998bh}
\begin{eqnarray}
	I^{(1)}_g &=&
		{} - \left(\frac{{\mu_r^2}}{-s-i\delta}\right)^\epsilon
		\frac{e^{\epsilon\gamma_E}}{\Gamma(1-\epsilon)}
		\frac{1}{\epsilon^2}
		\Big[
			C_A + 2\epsilon\beta_0
		\Big]\,,
                           \label{eq::I1}
\end{eqnarray}
with
\begin{eqnarray}
\label{eq::b0Hgdef}
  \beta_0 &=& \frac{1}{4}\left(\frac{11}{3}C_A - \frac{4}{3} T n_l \right)
  \,.
\end{eqnarray}

In addition to the form factors, in {\tt ggxy} we also implement
the virtual finite NLO corrections in the form
(see, e.g., Refs.~\cite{Heinrich:2017kxx,Grober:2017uho})
\begin{eqnarray}
  \widetilde{\mathcal{V}}_{\textnormal{fin}} &=&
  \frac{\alpha_s^2\left(\mu_r\right)}{16\pi^2}\frac{G_F^2 s^2}{64}
  \left[   C + 2\left( {\tilde{F}_1^{(0)*}}
    \tilde{F}_1^{(1)} + {\tilde{F}_2^{(0)*}} \tilde{F}_2^{(1)} +
    \tilde{F}_1^{(0)} {\tilde{F}_1^{(1)*}}+\tilde{F}_2^{(0)}
          {\tilde{F}_2^{(1)*}} \right) \right] \,,
      \label{eq::Vtil}
\end{eqnarray}
where $\tilde{F}^{(i)} = F^{{\rm fin},(i)}(\mu_r^2=-s)$ and
\begin{eqnarray}
  C &=& \left( \left|\tilde{F}_1^{(0)}\right|^2+\left|\tilde{F}_2^{(0)}\right|^2\right)
        \left(
        C_A\pi^2 - C_A\log^2\frac{\mu_r^2}{s}
        \right)
        \,,
\end{eqnarray}
with $\mu_r$ being the renormalization scale which is also present at the hadronic level.
Here, $\alpha_s$ corresponds to the five-flavour strong coupling constant. 
Furthermore, we introduce
\begin{eqnarray}
  \mathcal{V}_{\textnormal{fin}} 
  &=& \frac{\widetilde{\mathcal{V}}_{\textnormal{fin}}}{\alpha_s^2(\mu_r)}
      \,.
      \label{eq::Vfin}
\end{eqnarray}
Details on the implementation can be found in Section~\ref{sub::vfin}.

In the remaining part of this section we comment on the implementation of the analytic 
expressions.
After expansion in $m_H^2$, the expansion of the form factors in $t$ is a simple Taylor expansion with
coefficients depending on $s/m_t^2$. We express the amplitude in terms of
48 master integrals which we compute with the help of the ``expand and
match'' method~\cite{Fael:2021kyg,Fael:2021xdp,Fael:2022miw}. This provides
results for each $\epsilon$ coefficient of each master integral as a power-log
expansion around properly chosen values for $s/m_t^2$. The combination of these
expansions provides results with a precision of $10$ or more digits over the
whole $\sqrt{s}$ range. In practice we parametrize the $\epsilon$ expansion of
each master integral in terms of coefficients, which depend on $s/m_t^2$.  We
insert the generic expansion into the amplitude and convert this expression
into {\tt C++}. 
For the numerical evaluation we provide routines which implement the results of the ``expand
and match'' results for the coefficients of the master integrals.
In order to make the routine more efficient, we do not implement all of the series expansions provided by the method, 
but approximate the results via Chebychev polynomials, see e.g.~Ref.~\cite{Press:2007ipz}.
In practice we use Chebychev approximations with 100 terms in the regions 
$50 \leq s/m_t^2 < 500$, 
$25 \leq s/m_t^2 < 50$, 
$8 \leq s/m_t^2 < 25$, 
$4.2 \leq s/m_t^2 < 8$, 
$2 \leq s/m_t^2 < 3.8$, 
$0.1 \leq s/m_t^2 < 2$.
To cover the regions around the 
singular points $s/m_t^2= \{ 0 , 4 , \infty \}$ we include generalized series expansions with $\{10, 10, 7\}$ expansion terms, respectively.
The implementation of the master integrals rather than the approximation of the amplitude itself provides the 
possibility to re-use the implementation also for other scattering processes in the future.
For numerical stability in the limit $s/m_t^2 \to \infty$, however, it was necessary
to insert the expansions of the master integrals and implement the expanded amplitudes in {\tt C++}.
We switch to these explicit expansions for $s/m_t^2 > 500$.

In the high-energy limit one encounters a complicated asymptotic expansion for
$|s|,\allowbreak|t|,\allowbreak|u| \gg m_t^2$ which is discussed in detail in
Refs.~\cite{Davies:2018ood,Davies:2018qvx,Mishima:2018olh}. The efficient use
of the differential equations for the master integrals enables us to compute
more than 100~expansion terms in the limit $m_t\to0$. For the numerical
evaluation of the form factors based on these expansions, for $p_T\lesssim 500$ GeV, it is necessary to
construct Pad\'e approximations~\cite{Davies:2020lpf}. In
Ref.~\cite{Davies:2023vmj} it has been shown that a deep expansion in $m_t$
leads to precise results even close to the top--anti-top quark threshold if
$p_T\gtrsim 150$~GeV.
We implement the Pad\'e approximation following Ref.~\cite{Press:2007ipz}.
To be self-contained, we provide some details here: 

In the high-energy expansion, our results are given as series expansions in the variable $x=m_t^2/s$:
\begin{align*} 
    F(x) &= \sum\limits_{k=0}^{N+M} c_k x^k\,,
\end{align*}
where the $c_k$ can still depend on $\log(x)$.
We wish to find the Pad\'e approximant 
\begin{align*} 
    R(x) &= \frac{\sum\limits_{i=0}^{M} a_i x^i}{1+\sum\limits_{j=1}^{N} b_j x^j} ~,
\end{align*}
which satisfies
\begin{align}
    \left. \frac{ {\rm d}^i}{ {\rm d} x^i} F(x) \right|_{x=0} &= \left. \frac{ {\rm d}^i}{ {\rm d} x^i} R(x) \right|_{x=0}
\end{align} 
with $0\leq i<N+M$.
This problem is equivalent to finding the solution to the following system of equations: 
\begin{align}
    \sum\limits_{m=1}^{N} b_m C_{N-m+k} &= -c_{N+k} ,& k&=1,...,N ~, \\ 
    \sum\limits_{m=0}^{k} b_{m} c_{k-m} &= a_k ,& k&=1,...,N~. 
\end{align}
In order to solve this equation we use 
the QR decomposition with the help 
of Householder transformations implemented in the \texttt{Eigen} library~\cite{eigen}.
We follow Ref.~\cite{Davies:2020lpf} and compute several Pad\'e approximations and combine 
them in a weighted way to obtain a central value and an uncertainty estimate of our procedure.
Numerical instabilities can show up when we compute Pad\'e approximations for 
rather small values of $p_T \lesssim 200\,\text{GeV}$, where the expansion coefficients become very large.
The instabilities can be identified due to a large Pad\'e uncertainty on the approximation in a region where
our procedure should still provide accurate results. 
Whenever we find these large uncertainty estimates we rerun the Pad\'e procedure in quadruple precision.
Although we have over 100~expansion coefficients at hand, we implement only the first 
48 terms in \texttt{ggxy}; higher expansion terms quickly become numerically
unstable in double precision.
We observe that this is sufficient to obtain precise numerical values down to 
$p_T \sim 175$ GeV, where we can switch to the small-$t$ expansion.
In practice, we interpolate between the small-$t$ and high-energy expansions in the region
200 GeV $< p_T < 220$ GeV.


\section{\label{sec::real}Real radiation}

The proper treatment of IR divergences at NLO in QCD is very well studied.
However, in order to motivate the implementations in
{\tt ggxy} we briefly repeat the main features.
 
The partonic NLO QCD cross section can be written schematically as 
\begin{equation}
\sigma_{ab}^{\rm NLO}=\int_{n+1}\dd \sigma_{ab}^{\rm R}+\int_n\left[\dd\sigma_{ab}^{\rm LO+V}+\dd\sigma_{ab}^{\rm C}\right],
\end{equation}
where the subscript on the integrals indicates the dimension of the phase-space integration, $\sigma_{ab}^{\rm R}$ is the partonic process with an additional parton, $\dd\sigma_{ab}^{\rm LO+V}$ is the combination of the LO cross section and the virtual NLO corrections, and $\dd\sigma_{ab}^{\rm C}$ is the counterterm coming from the redefinition of PDFs due to the absorption of initial-state collinear singularities. While the partonic NLO cross section is IR finite, $\dd\sigma_{ab}^{\rm V}$ and $\dd\sigma_{ab}^{\rm C}$ contain explicit IR $\epsilon$ poles and further IR singularities arise in the first term after the phase-space integration of the unresolved parton in $d=4-2\epsilon$ dimensions. In order to make a phase-space integration in $d=4$ dimensions possible, we use the Catani-Seymour dipole subtraction scheme \cite{Catani:1996vz}, where the partonic cross section is rewritten as
\begin{equation}
\sigma_{ab}^{\rm NLO}=\int_{n+1}\left[\sigma_{ab}^{\rm R}-\sigma_{ab}^{\rm A}\right]+\int_n\left[\dd\sigma_{ab}^{\rm LO+V}+\dd\sigma_{ab}^{\rm C}+\int_1\dd\sigma_{ab}^{\rm A}\right],
\label{eq::sig_nlo}
\end{equation}
where a new subtraction term $\dd\sigma_{ab}^{\rm A}$ is introduced which mimics $\sigma_{ab}^{\rm R}$ in all IR limits and makes the integrand of the first term IR finite. The same term is added back to the second integrand and the phase-space integration over the unresolved parton leads to explicit IR $\epsilon$ poles which cancel exactly those from $\dd\sigma_{ab}^{\rm LO+V}$ and $\dd\sigma_{ab}^{\rm C}$. In summary, both terms are separately IR finite and can be safely integrated over the phase space in $d=4$ dimensions. 

In the Catani-Seymour subtraction scheme the last two terms in Eq.~(\ref{eq::sig_nlo})
are rewritten in terms of the so-called integrated dipole operators as
\begin{equation}
\int_n\left[\dd\sigma_{ab}^{\rm C}+\int_1\dd\sigma_{ab}^{\rm A}\right]=\int_n\left[\mathbf{I}({ \mu_r^2})\otimes\dd\sigma_{ab}^{\rm LO} + \sum_{a',b'}\int_0^1\dd x\, \mathbf{KP}_{ab,a'b'}(x,\mu_f^2)\otimes \,\dd\sigma_{a'b'}^{\rm LO}(x) \right],
\end{equation}
where the symbol $\otimes$ denotes colour correlations and the dependence on the IR and factorization scales is made explicit. The explicit definitions of the operators can be found in Ref.~\cite{Catani:1996vz}. The $\mathbf{KP}_{ab,a'b'}$ operator is further convoluted with the LO cross section and contains non-diagonal terms in flavour space with respect to the initial-state partons. Since at NLO we can have at most one $1\to2$ splitting, all terms in the $\mathbf{KP}_{ab,a'b'}$ are proportional to either $\delta_{aa'}$ or $\delta_{bb'}$. 

For the $gg\to HH$ process we have implemented the two initial-state dipoles with initial-state spectators
corresponding to the $g\to gg$ and $g\to q\bar{q}$ splittings. In addition, we have implemented the phase-space 
restriction on the subtraction terms~\cite{Nagy:1998bb,Nagy:2003tz} parametrized by $\alpha_{\rm CS}$. This parameter can be used to cross-check the calculation since the sum of all parts is independent of this artificial parameter, where the case $\alpha_{\rm CS}=1$ corresponds to no phase-space restriction. In {\tt ggxy} we set $\alpha_{\rm CS}=0.1$.
Following Ref.~\cite{Czakon:2009ss}, the parameter $\alpha_{\rm CS}$ is also used to parametrize a technical cut to avoid numerical instabilities due to large cancellations between the real emission contribution and the subtraction terms. We have cross-checked our implementation of the Catani-Seymour subtraction scheme against the program {\tt Helac-Dipoles} \cite{Czakon:2009ss} for single phase-space points and after phase-space integration. The finite part of the real corrections can be safely combined with the finite virtual correction $\widetilde{\mathcal{V}}_{\textnormal{fin}}$, defined in Eq.~\eqref{eq::Vtil}, where the latter contribution has to be multiplied by a factor of $\frac{\alpha_s}{2\pi}$.

For the real corrections to Higgs boson pair production, which have already been calculated in Refs.~\cite{Maltoni:2014eza,Borowka:2016ehy,Borowka:2016ypz,Baglio:2018lrj,Bagnaschi:2023rbx,Campbell:2024tqg}, the following sub-processes have to be taken into account
\begin{equation}
\begin{array}{clllcll}
gg &\to& HHg,
&\,&
gq/qg&\to& HHq\,,\\
q\bar{q}/\bar{q}q &\to& HHg,
&\,&
g\bar{q}/\bar{q}g&\to& HH\bar{q}.
\end{array}
\end{equation}
The corresponding one-loop amplitudes, as well as the spin-correlated squared matrix element of $gg\to HH$ required for the subtraction terms, are calculated with {\tt Recola}~\cite{Actis:2012qn,Actis:2016mpe} where the one-loop matrix elements are written in terms of tensor coefficients $c^{(t)}_{\mu_1...\mu_{r_t}}$ and tensor integrals $T^{\mu_1...\mu_{r_t}}_{(t)}$ as
\begin{equation}
\mathcal{M}_{\rm 1-loop}=\sum_t c^{(t)}_{\mu_1...\mu_{r_t}} T^{\mu_1...\mu_{r_t}}_{(t)}.
\end{equation}
The tensor integrals are defined as
\begin{equation}
T_{(t)}^{\mu_1...\mu_{r_t}}=\frac{(2\pi\mu)^{4-D}}{i\pi^2}\int \dd^Dq \frac{q^{\mu_1}...q^{\mu_{r_t}}}{D_0^{(t)}...D_{k_t}^{(t)}},
\end{equation}
where $k_t$ is the number of propagators and $r_t$ the rank of the tensor integral. The denominators are given by
\begin{equation}
D_i^{(t)}=\left(q+p_i^{(t)}\right)^2-\left(m_i^{(t)}\right)^2,
\end{equation}
with $p_0^{(t)}=0$. 

The tensor coefficients are calculated in a recursive approach in {\tt Recola} and the tensor integrals are calculated with {\tt Collier} \cite{Denner:2016kdg}, which performs an uncertainty estimation on the precision of the tensor integrals. If we encounter a phase-space point that leads to tensor integrals that are marked as unstable by {\tt Collier}, we perform an alternative reduction to scalar integrals using the OPP reduction technique \cite{Ossola:2006us} with the program {\tt CutTools} \cite{Ossola:2007ax} by using the interface with {\tt Recola} of Ref.~\cite{Stremmer:2024ecl}. In this case, we construct the numerator of the one-loop integrals using the tensor coefficients calculated by {\tt Recola} (in double precision) and multiply it with the tensor $q^{\mu_1}...q^{\mu_{r_t}}$. The reduction as well as the calculation of the scalar integrals with {\tt OneLOop} \cite{vanHameren:2010cp} is done in quadruple precision.

The rescaling of the Higgs boson self coupling by $\kappa_\lambda=\lambda_3/\lambda_3^{\rm SM}$ in {\tt Recola} is performed with the already built-in function to rescale a specific tree-level vertex, which is sufficient for our purpose. On the other hand, {\tt Recola} does not support modifications of numerical input parameters after the process initialization, such as the top-quark mass, which would be required in the ${\rm \overline{MS}}$ top-quark mass scheme when using a dynamical scale definition for $\mu_t$. A similar problem has was already been encountered in Ref.~\cite{Degrassi:2022mro} when using {\tt Recola~2}~\cite{Denner:2017wsf}, where the authors reinitialized this program for each phase-space point, leading to an increase in computation time by a factor of $5$. Instead, we have implemented in {\tt Recola} the possibility to update the top-quark mass even after the initialization phase, following the same approach that is already used in {\tt Recola} to update the UV counterterms. In {\tt ggxy} this option is only enabled if the ${\rm \overline{MS}}$ top-quark mass scheme is used and results in only a moderate increase of computation time of about $10\%$ of the matrix elements. Because of this modification in the {\tt Recola} version included in \texttt{ggxy}, it is not straightforward to replace it with a different version.\footnote{Currently, all modifications necessary for this can be tracked by searching the tag {\tt dynamic_params} in the source code.}

As an additional cross check we have computed analytic results for the one-loop helicity amplitudes of $gg\to HHg$ in terms of scalar integrals that are evaluated with {\tt CutTools} and {\tt OneLOop}. For the computation of the helicity amplitudes we have generated all Feynman diagrams with {\tt qgraf} \cite{Nogueira:1991ex}, which are mapped onto different topologies and converted to {\tt FORM} \cite{Ruijl:2017dtg} notation with 
{\tt tapir}~\cite{Gerlach:2022qnc}
and 
{\tt exp}~\cite{Harlander:1998cmq,Seidensticker:1999bb}. The computation of the diagrams is then performed with the in-house code {\tt calc} and the reduction to master integrals is carried out with {\tt Kira} \cite{Maierhofer:2017gsa,Klappert:2020nbg}. We find good agreement between the calculation with {\tt Recola} and the analytic results. However, in the latter case the coefficients in front of the scalar integrals turn out to be not sufficiently numerically stable over the whole phase space.
For this reason we use the approach based on
{\tt Recola} as our default option.


\section{\label{sec::ggxy}Using {\tt ggxy}}

\subsection{Installation and structure}

{\tt ggxy} can be downloaded or cloned from the
repository hosted at \url{https://gitlab.com/ggxy/ggxy-release}.
It contains the files and directories
\begin{verbatim}
CMakeLists.txt    README.md    example-build.sh    examples/
include/          lib_ext/     src/
\end{verbatim}
where \verb|CMakeLists.txt| is the \verb|CMake| configuration file of {\tt ggxy}, \verb|README.md| contains useful information, \verb|example-build.sh| is an example script to build 
{\tt ggxy} using \verb|CMake|. The directory \verb|examples/| contains two subdirectories 
\verb|gghh-FF/| and \verb|gghh-nlo/| which contain examples 
for the usage of {\tt ggxy} 
to calculate the two form factors as well as $\mathcal{V}_{\textnormal{fin}}$, and the 
calculation of integrated and differential hadronic cross sections at LO and NLO QCD. The 
examples are discussed in more detail in the next section. The directories \verb|src/| and 
\verb|include/| contain the source code and the corresponding header files of {\tt ggxy}, 
respectively. For convenience the source code of the following external libraries is located 
in the directory \verb|lib_ext|:
\begin{itemize}
\item {\tt avhlib}~\cite{vanHameren:2007pt,vanHameren:2010gg}: Phase-space generation.
\item {\tt Collier}~\cite{Denner:2016kdg}: Numerical evaluation of one-loop functions for real radiation.
\item {\tt CRunDec}~\cite{Herren:2017osy}: Running and decoupling for $\alpha_s$ and the top quark mass.
\item {\tt CutTools}~\cite{Ossola:2007ax}: Fall-back option for reduction to scalar integrals.
\item {\tt OneLOop}~\cite{vanHameren:2010cp}:
Fall-back option for numerical evaluation of one-loop functions for real radiation.
\item {\tt Recola}~\cite{Actis:2016mpe}: Generation of amplitudes for real corrections.
\end{itemize}
They are directly compiled together and correctly linked with {\tt ggxy} by \verb|CMake|. In addition, it is required that the following libraries are already installed:
\begin{itemize}
\item {\tt boost}~\cite{boost} and {\tt eigen}~\cite{eigen}: {\tt C++} libraries with convenient containers and functions, in particular in the context of linear algebra.
\item {\tt LHAPDF}~\cite{Buckley:2014ana,LHAPDF}: Provides parton distribution functions.
\end{itemize}

In order to evaluate polylogarithmic functions we have included the code from the ancillary files of
Ref.~\cite{Frellesvig:2016ske} directly into \texttt{ggxy}, which can be found in \verb|src/lievaluate/|.

For the installation of {\tt ggxy} it is sufficient to provide only the path to the {\tt LHAPDF} directory which contains the directories \verb|include| and \verb|lib|/\verb|lib64|, in the variable \verb|LHAPDFPATH| in example build script. In addition, it is possible to compile {\tt ggxy} only for the evaluation of the form factors by setting \verb|onlyFF=On|
in \verb|example-build.sh|. In this case the path to {\tt LHAPDF} is ignored and only {\tt CRunDec} from the external libraries is compiled together with {\tt ggxy}. 
With \verb|onlyFF=On| the examples in \verb|examples/gghh-FF/| 
can still be compiled whereas the examples in  \verb|examples/gghh-nlo/| require \verb|onlyFF=Off|. Further details about the compilation can be found in \verb|README.md|.

By running the installation script all external libraries of \verb|lib_ext/| and {\tt ggxy} are built in the directory \verb|example-build/|. In addition, a second directory is created, \verb|example-install/|, that contains the subdirectory \verb|include/| with all header and module files of the external libraries and {\tt ggxy}. The shared-libraries of all external libraries and {\tt ggxy} are placed in the directory \verb|example-install/lib/|, so that the content of \verb|example-install/| is sufficient to link {\tt ggxy} with other programs.

\subsection{Partonic form factors}
\label{sec:ff-functions}

The elementary building blocks implemented at the partonic level are
functions for the form factors $F_1$ and $F_2$ at one- and two-loop order in QCD.
{\tt ggxy} allows for an easy access to the 
finite parts of the form factors as defined in Eq.~(\ref{eq::FF_IR}).
The function prototype looks as follows
\begin{lstlisting}
complex<double> gghhFF(int loops, int ff, double s, double t,
                       double mhs, double mts, 
                       double murs = gghhFFmursDefault,
                       double muts = 0.0, unsigned scheme = 0, 
                       double kappa_lam = 1.0, double dTriCoeff = 1.0);
\end{lstlisting}
where the parameters are defined as follows,
\begin{itemize}
    \item \verb|loops|: QCD loop order, 1 or 2
    \item \verb|ff|: choice of form factor, 1 or 2
    \item \verb|s,t|: Mandelstam variables
    \item \verb|mhs|: squared Higgs boson mass, $m_H^2$
    \item \verb|mts|: squared top quark mass, $m_t^2$
    \item \verb|murs|: squared renormalization scale $\mu_r^2$ 
    \item \verb|muts|: squared renormalization scale for the $\overline{\rm MS}$ top quark mass $\mu_t^2$
    \item \verb|scheme|: choice of renormalization scheme for top quark mass, 0 (OS) or 1 ($\overline{\rm MS}$). 
    \item \verb|kappa_lam|: corresponds to $\kappa_\lambda$
    \item \verb|dTriCoeff|: additional coefficient for the double-triangle contribution, see Eq.~(\ref{eq::F_12}); can be used, for e.g., to switch on and off the double-triangle contribution
\end{itemize}
The default value of \verb|murs| is set to the pre-processor variable \verb|gghhFFmursDefault| which stands for the choice $\mu_r^2=-s$.
At one-loop
order the exact results~\cite{Glover:1987nx,Plehn:1996wb} are implemented.\footnote{At the border of the phase space we switch to expansions which provide more stable results.}
At two-loop order, depending on the numerical value of $p_T$, routines for
either the small-$t$ or the high-energy expansions~\cite{Davies:2023vmj} are called.

Using the function \verb|gghhFF()| it is straightforward to construct 
the quantity ${\cal V}_{\rm fin}$ in Eq.~(\ref{eq::Vfin})
using the on-shell or $\overline{\rm MS}$ scheme for different
choices of $\mu_t$.
For convenience {\tt ggxy} provides the function
\begin{lstlisting}
double gghh2lVfin(double s, double t, double mhs, double mts, double GF,
                  double murs = gghhFFmursDefault, double muts = 0.0,
                  unsigned scheme = 0, double kappa_lam = 1.0);  
\end{lstlisting}
where the meaning of the parameters is as above with the additional parameter \verb|GF| for $G_F$.
In this case, the pre-processor variable \verb|gghhFFmursDefault| stands for $\mu_r^2=s/4$.
Note that this is in contrast to the default value of \verb|murs| = $\mu_r^2=-s$ used for the form factors. \verb|gghh2lVfin| only accepts positive values for \verb|murs|.

In addition, we provide the function
\begin{lstlisting}
double gghh1l(double s, double t, double mhs, double mts,
              double GF, double kappa_lam = 1.0);
\end{lstlisting}
which can be used to calculate the LO squared matrix element with a factor of $\alpha_s^2$ is factored out, and the function
\begin{lstlisting}
vector<double> gghh2l(double s, double t, double mhs, double mts,
                      double GF, double murs = gghhFFmursDefault,
                      double muts = 0.0, unsigned scheme = 0,
                      double kappa_lam = 1.0);
\end{lstlisting}
which returns the LO squared matrix element as the first vector element and ${\cal V}_{\rm fin}$ as the second.

\subsection{Functions for hadronic cross sections}

The calculation of hadronic total and differential cross sections is managed by the class \verb|mc_gen|
which has the following functionality:
\begin{itemize}
    \item Set input parameters ($m_t$, $m_H$, top-quark mass scheme, renormalization and factorization scales, PDF set, \ldots).
    \item Integrate all contributions needed for NLO predictions together, or each contribution individually.
    \item Optimize phase-space integration.
    \item Perform Monte-Carlo integration.
    \item Fill results into histograms and write final results to disk.
\end{itemize}
The Monte-Carlo integration is constructed based on several 
integration channels for the different subprocesses in the real 
corrections and of the individual contributions with a Born-like 
phase space. The individual weights of these channels are optimized 
during the optimization phase of the program following the approach 
of Ref.~\cite{Kleiss:1994qy}. The phase spaces of Born-like 
contributions and of the real corrections are generated with {\tt Kaleu} \cite{vanHameren:2007pt,vanHameren:2010gg} as part of the {\tt avhlib} library, which is a multi-channel phase-space generator that performs further 
optimizations on-the-fly. 

In the following we provide a brief summary
of the functions that can be called; a
detailed example of the class is 
given in the file \verb|examples/gghh-nlo/nlo-gghh.cpp|. 
The \verb|mc_gen| class is called as
\begin{lstlisting}
mc_gen gen = mc_gen(int seed, double ss);
\end{lstlisting}
where the first parameter is the seed for the initialization of the random numbers and the second parameter is the hadronic centre-of-mass energy $\sqrt{s}$ in units of GeV. The structure of this class is process independent. However, currently we offer only 
a configuration of \verb|mc_gen| for the production of two Higgs bosons at the LHC that is called with
\begin{lstlisting}
configure_mc_gghh(string& mode, mc_gen& gen);
\end{lstlisting}
The first parameter defines the contribution that should be integrated over the phase space and can be one of the following keywords:
\begin{itemize}
    \item \verb|lo|: LO cross section
    \item \verb|nlo|: NLO cross section
    \item \verb|V|: Virtual corrections given by $\frac{\alpha_s}{2\pi}\widetilde{\mathcal{V}}_{\textnormal{fin}}$
    \item \verb|I|: $\mathbf{I}$ operator of Catani-Seymour subtraction scheme
    \item \verb|KP|: Sum of $\mathbf{K}$ and $\mathbf{P}$ operator of Catani-Seymour subtraction scheme
    \item \verb|RS|: Real subtracted contribution using Catani-Seymour subtraction scheme
    \item \verb|LOVDIP|: Sum of \verb|lo|, \verb|V|, \verb|I| and \verb|KP|.
\end{itemize}
Note that \verb|nlo| corresponds to the sum of \verb|LOVDIP| and \verb|RS|.

The parameters of the process should be modified by one of the following class methods
\begin{lstlisting}
void set_mass_top(double mtop);
void set_mass_higgs(double mhiggs);
void set_gfermi(double gfermi);
\end{lstlisting}
where the default values can be found in \verb|src/tools/params.cpp|. In addition, it is possible to rescale the trilinear self-coupling of the Higgs boson by $\kappa_\lambda$ with the function
\begin{lstlisting}
void set_kappa_lam(double kappa_lam);
\end{lstlisting}
The top-quark mass scheme can be set with the function
\begin{lstlisting}
void set_mtscheme(unsigned mtscheme      , int crd_runLoops_as  = 5,
                  int crd_runLoops_mt = 5, int crd_convLoops_mt = 4);
\end{lstlisting}
where again \verb|mtscheme=0| corresponds to the on-shell and \verb|mtscheme=1| to the ${\rm \overline{MS}}$ scheme. The last three parameters of the function \verb|set_mtscheme| control the loop order for the running of the strong coupling constant and top-quark mass, and the loop order for the conversion from the on-shell to the $\overline{\rm MS}$ top-quark mass. The conversion and the running are performed with \verb|CRunDec| and the default values of the parameters are set to highest orders available. The input values $\alpha_s(m_Z)$ and $m_Z$ for the running are read from the PDF set. The strong coupling constant, which appears explicitly in the matrix element, is always calculated with \verb|LHAPDF|, where the parameters are controlled by the given PDF set. Thus, the last two parameters are always ignored in the on-shell scheme.

The generator is able
to produce results for different 
values of the renormalization and factorization scales in a single 
run. In particular, it is possible to perform the usual
$7$-point scale variation to 
estimate theoretical uncertainties, where the central values ($\mu_{r,0}$, $\mu_{f,0}$) of the 
renormalization and factorization scales are varied as follows:
\begin{equation}
\left( {\frac{\mu_r}{\mu_{r,0}}} ,\frac{\mu_f}{\mu_{f,0}} \right)\in \left\{ (1,1),(0.5,0.5),(2,2),(2,1),(0.5,1),(1,2),(1,0.5) \right\}.
\label{eq::7point}
\end{equation}
The names of the corresponding scale definitions, the information
about the variation, and the PDF set is either initialized with
\begin{lstlisting}
void set_scales_pdfs(vector<string>& scale_names, 
                     int gridtype, 
                     string pdf_name);
\end{lstlisting}
or
\begin{lstlisting}
void set_scales_pdfs(vector<string>& scale_names, 
                     vector<vector<double>> igrid, 
                     string pdf_name);
\end{lstlisting}
where the first input parameter is a vector with the names of the 
scales. They are only used for bookkeeping purposes and appear as a label
in the histogram files which are generated by {\tt ggxy}.
The actual definition of the scales happens in the function \verb|set_scale| (see also below)
where vectors for the renormalization and factorization scales, with the same length
as the vector containing the names of the 
scales, are introduced.
The last input parameter, \verb|set_scales_pdfs|, is the name of the PDF set as 
defined in \verb|LHAPDF|. The second input parameter is either an 
integer number to use a predefined grid, where \verb|gridtype=0| 
corresponds to the case where only the central value of the scale is 
calculated. With \verb|gridtype=1| the $7$-point scale variation 
as given in Eq.~(\ref{eq::7point})
is performed. Alternatively, a matrix that contains the 
information about the scale variations can be provided.
For example, the $7$-point scale variation of Eq.~(\ref{eq::7point}) can also be calculated with 
\begin{small}
\begin{lstlisting}
igrid={{1.0,1.0},{0.5,0.5},{2.0,2.0},{2.0,1.0},{0.5,1.0},{1.0,2.0},{1.0,0.5}}
\end{lstlisting}
\end{small}
The renormalization and factorization scales are then defined with
\begin{lstlisting}
void set_scalefunc(scalefunc set_scale);
\end{lstlisting}
where \verb|scalefunc| is a type definition of a function (or pointer to a function) with the following prototype
\begin{lstlisting}
void set_scale(vector<int>& iproc, 
               vector<lorentz_vec>& p, 
               params& pars, 
               vector<double>& muR, vector<double>& muF, 
               double& mut);
\end{lstlisting}
Here \verb|iproc| is a vector containing the identification of the 
particles of the process following the conventions of the 
PDG~\cite{ParticleDataGroup:2022pth} and \verb|p| is a vector 
containing the corresponding four momenta in the center-of-mass 
system. The parameter \verb|pars| is a class containing all 
numerical parameters, so that the top-quark or Higgs boson mass can be obtained 
by \verb|pars.mt| or \verb|pars.mh|, respectively. The different 
scale settings for the renormalization and factorization scales are 
defined in the vectors \verb|muR| and \verb|muF|, where the length of 
these vectors is equal to the length of the vector \verb|scale_names| 
which has been used in the function \verb|set_scales_pdfs| to 
initialize the scale and PDF settings. In the case of the 
$\overline{\rm MS}$ top-quark mass scheme the corresponding scale 
$\mu_t$ has to be passed as the parameter \verb|mut|. For convenience the 
top-quark mass $\overline{m}_t(\overline{m}_t)$ is stored in \verb|pars.mtmt|. 

Histograms can be defined and filled with 
the functions
\begin{lstlisting}
void set_hinit(histo_init hinit);
\end{lstlisting}
and
\begin{lstlisting}
void set_hfill(histo_fill hfill);
\end{lstlisting}
where \verb|histo_init| and \verb|histo_fill| are type definitions to functions with the following prototypes
\begin{lstlisting}
void hinit(histo& hlist);
\end{lstlisting}
and
\begin{lstlisting}
void hfill(histo& hlist, 
           vector<int>& iproc,
           vector<lorentz_vec>& plab,
           params& pars);
\end{lstlisting}
The histograms can then be initialized in the function \verb|hinit| with the following class method of \verb|hlist|
\begin{lstlisting}
void add(string& name, int bn, 
         double start, double end);
\end{lstlisting}
where the first input parameter is the observable name, \verb|bn| is the number of bins and the variables \verb|start| and \verb|end| define the range of the histogram. Alternatively, it is possible to add histograms with
\begin{lstlisting}
void add(string& name, int bn, 
         vector<double>& bins);
\end{lstlisting}
where in this case the vector \verb|bins| should contain the edges of the histograms so that this function can be used to create histograms with unequal bin sizes. The observables should then be constructed in the function \verb|hfill| where the variables \verb|iproc| and \verb|pars| are the same as in the function \verb|set_scale| and \verb|plab| are the particle momenta in the laboratory frame. The histograms should then be filled in this function with
\begin{lstlisting}
void fill(int ih, double val);
\end{lstlisting}
which is a class method of \verb|hlist|. The variable \verb|ih| is used to identify a histogram defined in \verb|hfill|, where the first histogram defined with \verb|add| has \verb|ih|$=0$, and the variable \verb|val| is the value of the observable for this event. Finally, it is possible to define a function for possible phase-space cuts
\begin{lstlisting}
bool event_cut(vector<int>& iproc, 
               vector<lorentz_vec>& plab, 
               params& pars);
\end{lstlisting}
where the input parameters are identical to \verb|hfill|. This function should return \verb|false| if the event should be rejected and otherwise \verb|true|. The function can then be given to the generator by using the class method
\begin{lstlisting}
void set_event_cut(cutfunc event_cut);
\end{lstlisting}
of \verb|mc_gen|.

The initialization of the generator is considered complete when the class method
\begin{lstlisting}
void finish_init();
\end{lstlisting}
is called, after which none of the functions above should be called and the optimization phase is activated. In this phase the weights of all integration channels and the phase-space generator {\tt Kaleu} are optimized. The generation of phase-space points is then achieved with the class method
\begin{lstlisting}
vector<double> integrate(int npT, int istep=10000);
\end{lstlisting}
where the variable \verb|npT| is the number of phase-space points that should be generated and \verb|istep| defines the number of phase-space points after which log information is printed. The return vector contains the cross section of the first scale definition and the corresponding MC uncertainty. The optimization phase is deactivated by calling the class method
\begin{lstlisting}
void set_phase_optim(bool phase_optim);
\end{lstlisting}
with \verb|phase_optim=false|, after which the histogram will be filled in the next call of \verb|integrate|. All MC integration weights including the histograms, can be reset with the class method
\begin{lstlisting}
void reset_weights();
\end{lstlisting}
of \verb|mc_gen|. Finally, the histograms can be saved to disk with
\begin{lstlisting}
void output(string outfile);
\end{lstlisting}
where \verb|outfile| is the output file.


\section{\label{sec::examples}Example results}

The example files to compute partonic and hadronic quantities are in 
the subdirectories
\verb|examples/gghh-FF/| and \verb|examples/gghh-nlo/|, respectively. 
They are compiled by executing the script \verb|build.sh|
in the corresponding directory.

\subsection{One- and two-loop form factors}
Using the functions \verb|gghh<n>lFF<i>|
it is straightforward to reproduce
numerical results for the form factors
present in the literature.
The example file \verb|examples/gghh-FF/ff.cpp|
shows how the data points for the (exact)
one- and two-loop curves in Fig.~3
of Ref.~\cite{Davies:2025ghl}
can be generated. In addition the
example shows how the corresponding data points
for a $\overline{\rm MS}$ top quark mass
can be generated.

\subsection{\label{sub::vfin}Virtual NLO corrections}

The ultraviolet renormalized and IR subtracted virtual
corrections $\mathcal{V}_{\textnormal{fin}}$ from Eq.~(\ref{eq::Vfin})
are implemented in the function \verb|gghh2lVfin()|, described above.
The example given in the file \verb|examples/gghh-FF/check-hhgrid.cpp|
calls this function for all 6320 data points
contained in the {\tt hhgrid}~\cite{hhgrid} interpolation grid
and evaluates them in less than 10~seconds;
this performance allows the use of \texttt{ggxy} for Monte-Carlo studies.
In principle the use of the grid for interpolation might be even faster, however it lacks the flexibility to change the parameters such as the masses of the top quark or the Higgs boson. 
Extending the grid, for example for a running top quark mass with an $m_{HH}$-dependent scale $\mu_t$ would require computational resources many orders of magnitude greater than our implementation.
All results from Ref.~\cite{hhgrid} are validated within the uncertainties.

\subsection{Hadronic cross sections}

The file \verb|examples/gghh-nlo/nlo-gghh.cpp| illustrates how
hadronic cross sections can be computed.
It contains auxiliary functions to define the renormalization and factorization scales and to initialize and fill the histograms.
In the main part of the program one first selects which part of the NLO corrections to compute and initializes the Monte-Carlo generator. Afterwards one specifies the input parameters using, e.g.
\begin{lstlisting}
  gen.set_mass_top(mt);
\end{lstlisting}
The renormalization scheme for the top quark mass is selected via
\begin{lstlisting}
  gen.set_mtscheme(mtscheme);
\end{lstlisting}
After initializing the histograms and specifying cuts\footnote{In our example no cuts are applied.}
it is possible to initiate the Monte-Carlo integration.
Finally the generated data are stored to disk.
They are used to obtain total and differential cross sections
which are discussed in the following subsections.

\subsubsection{Total hadronic cross section}

\begin{table}[t]
    \centering
    \renewcommand{\arraystretch}{1.2}
    \begin{tabular}{cc@{\hskip 10mm}l@{\hskip 10mm}l@{\hskip 10mm}}
        \hline
        $\sqrt{s}$&
        &{ \tt ggxy}&Ref.~\cite{Borowka:2016ypz}  \\
        \hline
        14~TeV & $\sigma^{\rm LO}$ [fb]& $19.848(4)^{+27.6\%}_{-20.5\%}$ &    $19.85^{+27.6\%}_{-20.5\%}$\\
        & $\sigma^{\rm NLO}$ [fb]& $32.92(2)^{+13.6\%}_{-12.6\%}$ & $32.91^{+13.6\%}_{-12.6\%}$\\
        \noalign{\smallskip}\hline\noalign{\smallskip}
        100~TeV & $\sigma^{\rm LO}$ [fb]& $731.2(2)^{+20.9\%}_{-15.9\%}$ &    $731.3^{+20.9\%}_{-15.9\%}$\\
        & $\sigma^{\rm NLO}$ [fb]& $1150(1)^{+10.8\%}_{-10.0\%}$ & $1149^{+10.8\%}_{-10.0\%}$\\        
        \noalign{\smallskip}\hline\noalign{\smallskip}
    \end{tabular}
    \caption{\label{tab::sig1}Comparison with results of Ref.~\cite{Borowka:2016ypz}
    for $\sqrt{s}=14$~TeV and $\sqrt{s}=100$~TeV.}
\end{table}

In the sample file \verb|examples/gghh-nlo/nlo-gghh.cpp| the total
hadronic cross section is computed using the input values
from Ref.~\cite{Borowka:2016ypz}. In particular, we use
the PDF set \verb|PDF4LHC15_nlo_100_pdfas|
which is obtained via the interface of LHAPDF~\cite{Buckley:2014ana}. The renormalization and factorization scales are set to a common scale $\mu_r=\mu_f=m_{HH}/2$. We reproduce
the central values and uncertainties based on the on-shell
top quark mass given
in Ref.~\cite{Borowka:2016ypz} both for $\sqrt{s}=14$~TeV and $\sqrt{s}=100$~TeV. The comparison is shown in Table~\ref{tab::sig1}. The results from {\tt ggxy} are produced by averaging the results from five different seeds. The relative uncertainty for each seed is about $0.2\%$ with a run time of about $30$ minutes on a single core on a AMD Ryzen Threadripper PRO 3955WX processor. An example script \verb|examples/gghh-nlo/run.sh| is provided to illustrate the usage of the sample file to compute cross sections with different seeds in parallel followed by a combination of the results using python scripts.

For illustration the total cross section using the $\overline{\rm MS}$ scheme can also be calculated with the example file by only setting \verb|mtscheme=1|, where we use $\mu_t=\overline{m}_t(\overline{m}_t)$. We obtain
\begin{eqnarray}
    \sigma_{\rm tot}(gg\to HH) &=& 31.63(2) ~\mbox{fb}
    \,.
\end{eqnarray}
The cross section is calculated again by averaging over five seeds which all have a relative uncertainty of less than $0.2\%$. The runtime increased slightly to about $35$ minutes.

\subsubsection{Differential distributions}

\begin{figure}[t]
  \begin{center}
  \begin{tabular}{cc}
     \includegraphics[width=0.45\textwidth]{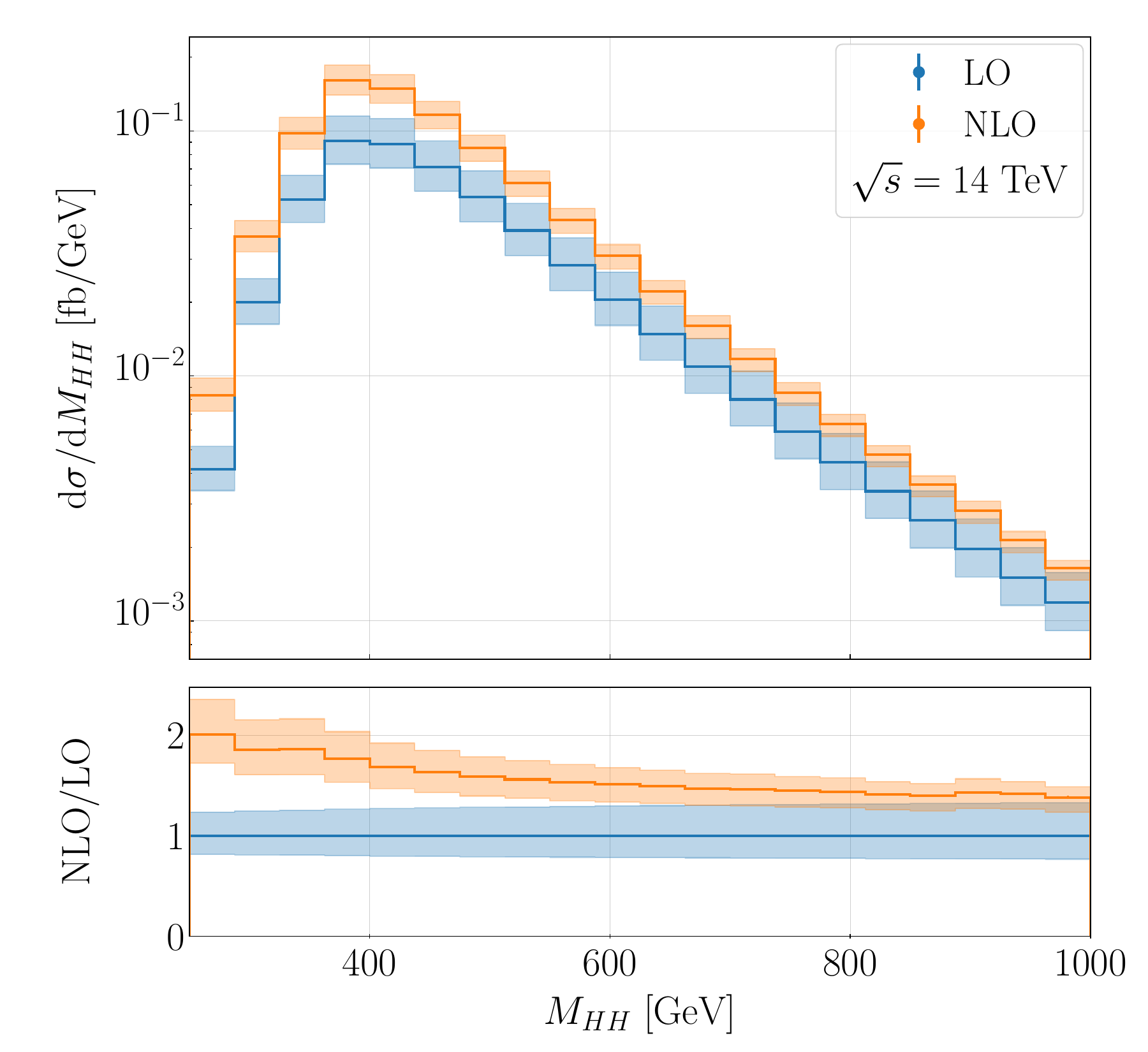} & 
     \includegraphics[width=0.45\textwidth]{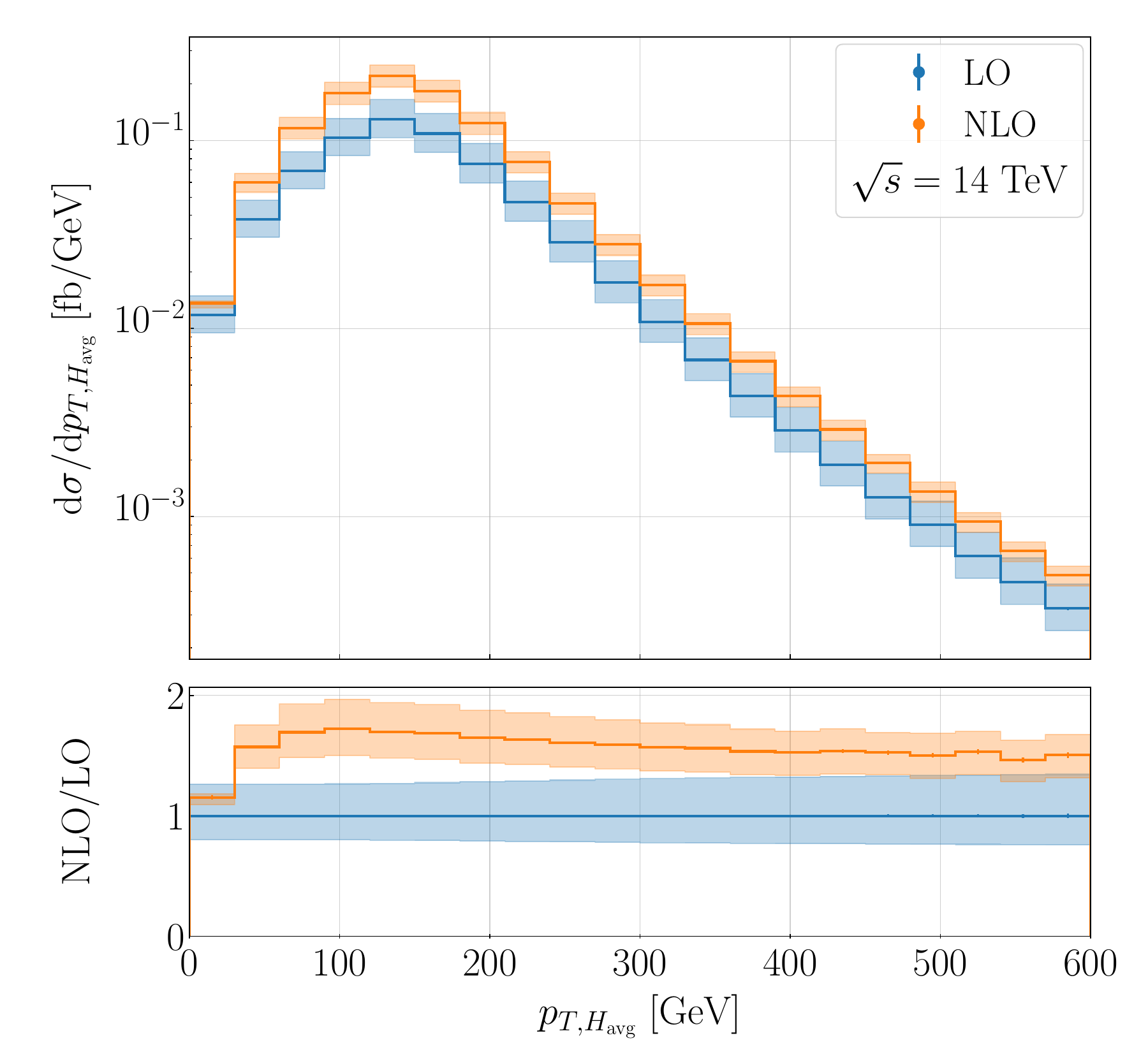}
  \end{tabular}
  \end{center}
  \caption{\label{fig::pT_mHH} LO and NLO QCD predictions for the invariant mass distribution of the di-Higgs system and the average of both transverse momenta of the two Higgs bosons. Theoretical uncertainties from scale variation are shown as bands and the statistical uncertainties from Monte Carlo integration are shown as vertical lines. Note that the latter are small and are only visible for larger values of $M_{HH}$ and $p_{T,H_{\rm avg}}$.}
\end{figure}

The data files generated in the runs for the total cross section in the above example also 
contain the data points for the differential cross section distributions. As an example in 
Fig.~\ref{fig::pT_mHH} the invariant mass distribution of the di-Higgs system and the average 
of both transverse momenta of the two Higgs bosons are presented at LO and NLO QCD.
These plots are generated by the \texttt{python} scripts provided in \verb|examples/gghh-nlo/|. By 
default, the example script considers five seeds. However, depending on the observable and 
the kinematic ranges, more seeds might be required to obtain smooth distributions.

\subsection{Variation of $\lambda$}

With the command
\begin{lstlisting}
gen.set_kappa_lam(kappa_lam);
\end{lstlisting}
the quantity $\kappa_\lambda$ can be modified.
\begin{table}[t]
    \centering
    \renewcommand{\arraystretch}{1.2}
    \begin{tabular}{cc@{\hskip 10mm}l@{\hskip 10mm}l@{\hskip 10mm}}
        \hline
         $\sqrt{s}$&$\kappa_{\lambda}$&$\sigma^{\rm NLO}_{\rm \tt ggxy}$ [fb] & $\sigma^{\rm NLO}_{\rm \mbox{\footnotesize \cite{Baglio:2020wgt}}}$ [fb]  \\
        \hline
        $13$ TeV & $-10.0$ & $1424.1(9)$ & $1438(1)$\\
        $13$ TeV & $-5.0$ & $509.4(3)$ & $512.8(3)$\\
        $13$ TeV & $-1.0$ & $113.53(7)$ & $113.66(7)$\\
        $13$ TeV & $0.0$ & $61.36(4)$ & $61.22(6)$\\
        $13$ TeV & $1.0$ & $27.72(2)$ & $27.73(7)$\\
        $13$ TeV & $2.0$ & $12.777(9)$ & $13.2(1)$\\
        $13$ TeV & $2.4$ & $12.035(8)$ & $12.7(1)$\\
        $13$ TeV & $3.0$ & $16.50(1)$ & $17.6(1)$\\
        $13$ TeV & $5.0$ & $80.00(5)$ & $83.2(3)$\\
        $13$ TeV & $10.0$ & $564.7(4)$ & $579(1)$\\
        $14$ TeV & $1.0$ & $32.79(2)$ & $32.81(7)$\\
        \noalign{\smallskip}\hline\noalign{\smallskip}
    \end{tabular}
    \caption{\label{tab::sig2}Comparison with results of Ref.~\cite{Baglio:2020wgt}.}
\end{table}
In Tab.~\ref{tab::sig2} we compare the results for 
the total cross section to
Refs.~\cite{Baglio:2020wgt} (see also Ref.~\cite{Baglio:2020ini})
for different choices of $\kappa_\lambda$.
We observe good agreement within the uncertainties 
for small values of $\kappa_\lambda$ (i.e., for $|\kappa_\lambda|\lesssim 1$). However for larger values
of $|\kappa_\lambda|$ significant differences are observed. Similar
deviations are also observed in Ref.~\cite{Bagnaschi:2023rbx}, see the discussion in Section 3.1 of that paper.

\begin{table}[t!]
    \centering
    \renewcommand{\arraystretch}{1.2}
    \begin{tabular}{cc@{\hskip 10mm}l@{\hskip 10mm}l@{\hskip 10mm}}
        \hline
         $\kappa_{\lambda}$&Top-mass scheme&$\sigma^{\rm NLO}_{\rm \tt ggxy}$ [fb] & $\sigma^{\rm NLO}_{\rm \mbox{\footnotesize \cite{Bagnaschi:2023rbx}}}$ [fb]  \\
        \hline
        $-0.6$ & on-shell & $100.34(6)^{+15.7\%}_{-13.6\%}$ & $100.77^{+15.8\%}_{-13.7\%}$\\
        $0.0$ & on-shell & $68.08(4)^{+15.1\%}_{-13.4\%}$ & $68.38^{+15.1\%}_{-13.4\%}$\\
        $1.0$ & on-shell & $30.83(2)^{+13.8\%}_{-12.7\%}$ & $30.93^{+13.7\%}_{-12.7\%}$\\
        $1.0$ & $\overline{\rm MS},\,\mu_t=\overline{m}_t(\overline{m}_t)$ & $29.78(2)^{+14.3\%}_{-13.0\%}$ & $29.78^{+14.3\%}_{-13.0\%}$\\
        $1.0$ & $\overline{\rm MS},\,\mu_t=M_{HH}/2$ & $28.79(2)^{+15.3\%}_{-13.5\%}$ & $28.90^{+15.2\%}_{-13.5\%}$\\
        $2.4$ & on-shell & $13.369(9)^{+14.7\%}_{-13.2\%}$ & $13.41^{+14.8\%}_{-13.1\%}$\\
        $6.6$ & on-shell & $203.4(1)^{+19.0\%}_{-15.1\%}$ & $203.91^{+19.0\%}_{-15.2\%}$\\
        \noalign{\smallskip}\hline\noalign{\smallskip}
    \end{tabular}
    \caption{\label{tab::sig3}Comparison with results of Ref.~\cite{Bagnaschi:2023rbx}.}
\end{table}

Finally, in Tab.~\ref{tab::sig3} we compare to the results
provided in Ref.~\cite{Bagnaschi:2023rbx} for different values of
$\kappa_\lambda$. 
After adopting their input parameters
i.e., $\sqrt{s}=13.6$~TeV, $M_t=172.5$~GeV, $m_H=125$~GeV
and the PDF set \verb|NNPDF31_nlo_as_0118| we observe good agreement,
however in Ref.~\cite{Bagnaschi:2023rbx} no Monte Carlo uncertainties are given
with which we could compare. 
We also agree with the results given in the revised version of Ref.~\cite{Buchalla:2018yce}, see 
also Ref.~\cite{Heinrich:2019bkc}.

\subsection{Top quark mass renormalization scheme dependence}

\begin{figure}[t]
  \begin{center}
  \begin{tabular}{cc}
     \includegraphics[width=0.45\textwidth]{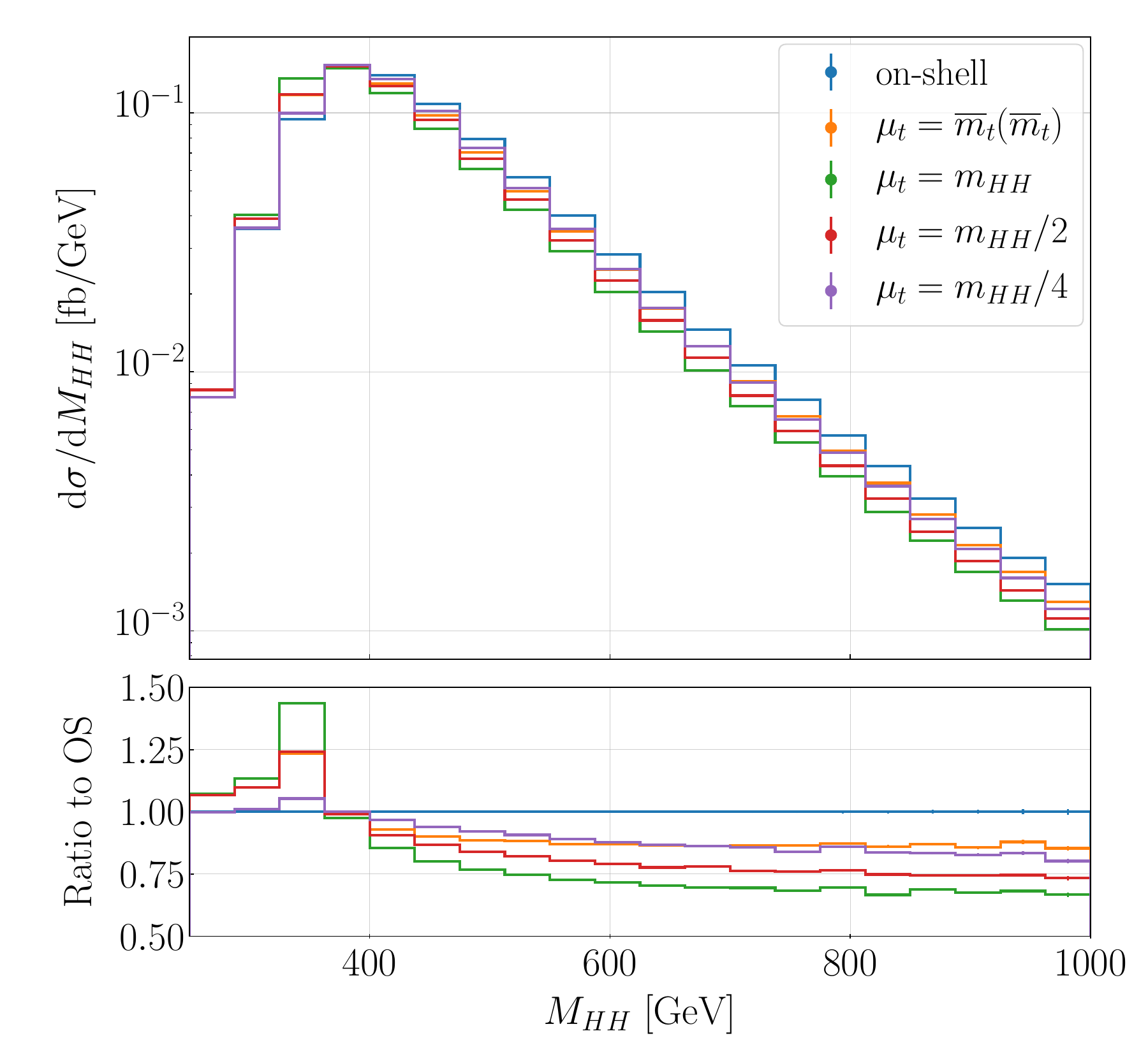}
     \includegraphics[width=0.45\textwidth]{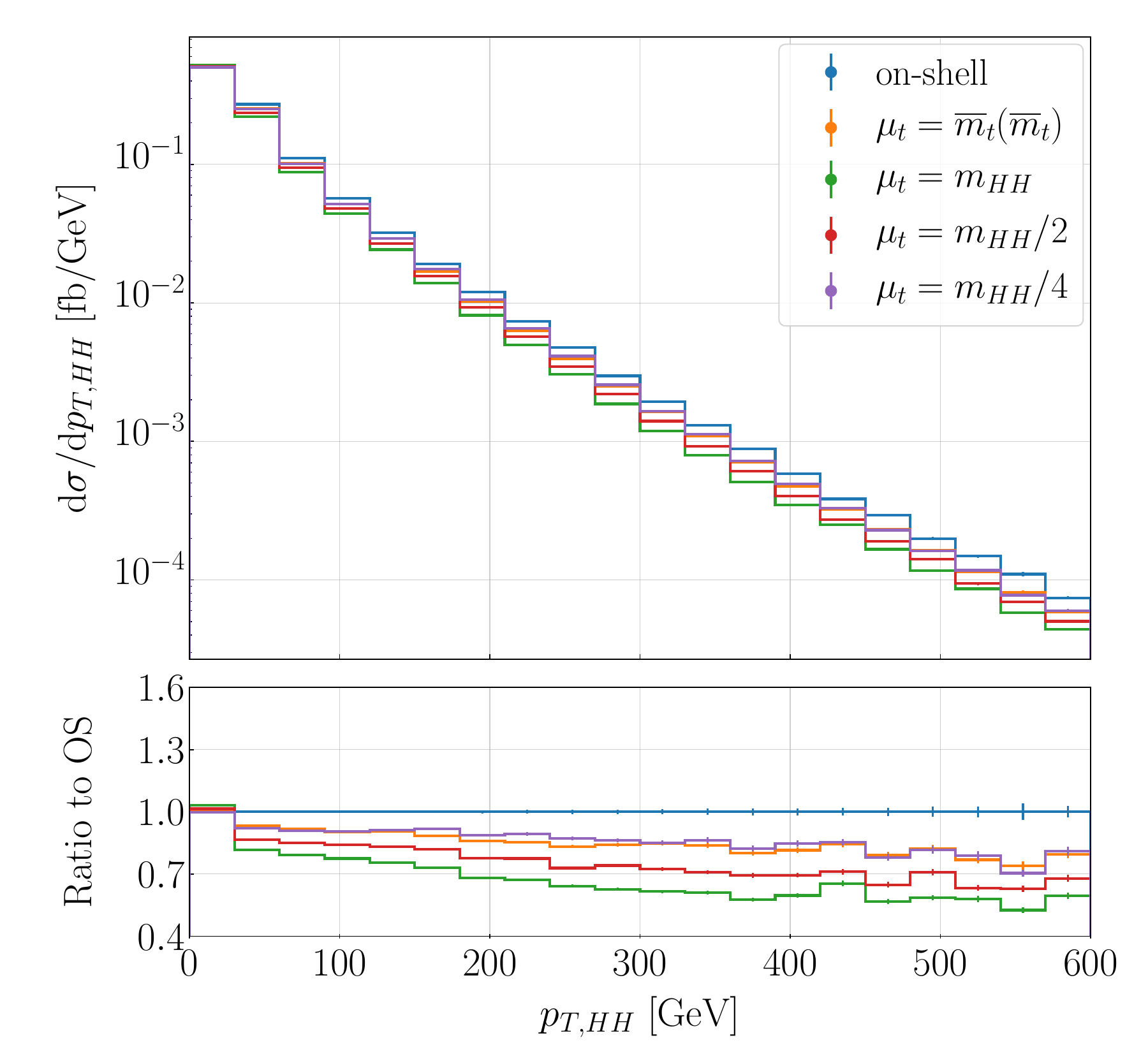}
  \end{tabular}
  \end{center}
  \caption{\label{fig::mtMSOS}
   NLO predictions for $m_{HH}$ and $p_{T,HH}$ distributions using the on-shell mass scheme as well as the $\overline{\rm MS}$ scheme for different settings of $\mu_t$. Input parameters are adapted from Ref.~\cite{Bagnaschi:2023rbx} with $\sqrt{s}=13.6$ TeV. The lower panels display the ratio to the on-shell predictions. Statistical uncertainties from Monte Carlo integration are shown as vertical lines.}
\end{figure}

The sample file \verb|examples/nlo-gghh.cpp| can also be used to study the dependence on the top quark mass renormalization scheme. We perform separate runs for
the on-shell (\verb|mtscheme=0|) and the $\overline{\rm MS}$ scheme (\verb|mtscheme=1|). The scale $\mu_t$ can be modified in the function \verb|set_scale|. For example, $\mu_t=\overline{m}_t(\overline{m}_t)$
corresponds to
\verb|mut = pars.mtmt| and $\mu_t=m_{HH}/2$ is obtained with
\verb|mut = phh.mass()/2.0|.
In Fig.~\ref{fig::mtMSOS} we show the differential distributions for the observables $m_{HH}$ and $p_{T,HH}$ in the on-shell and the $\overline{\rm MS}$ scheme for different choices of $\mu_t$. The lower panels display the ratio to the on-shell predictions.
We reproduce the results from Refs.~\cite{Baglio:2020ini,Bagnaschi:2023rbx}.


\section{\label{sec::concl}Conclusions and outlook}

In this paper we present the fast and flexible library {\tt ggxy}
which can be used to compute partonic and hadronic
quantities related to the loop-induced gluon-initiated processes.
In Version~1 we implement the functionality which allows
the computation of NLO QCD corrections to
Higgs boson pair production. Example files are
provided which demonstrate how to compute LO and NLO
corrections to the form factors, NLO virtual corrections,
total cross sections and distributions. All
results can be obtained for on-shell or $\overline{\rm MS}$ top quark masses. Furthermore, it is possible
to modify the self-coupling of the Higgs boson $\lambda$. The typical runtime for partonic quantities
is a few milliseconds and for hadronic quantities minutes to hours.
The high degree of flexibility and the
low runtime makes {\tt ggxy} attractive for use as an amplitude library for
parton shower programs such as, e.g., {\tt POWHEG}~\cite{Alioli:2010xd}.

It is straightforward to extend {\tt ggxy} in various
directions. NLO QCD corrections to processes such as top-quark mediated $gg\to ZZ$ or $gg\to ZH$
can be implemented in complete analogy to $gg\to HH$ and will be made available in a future version.
In future versions we additionally plan to implement NNLO QCD
and NLO electroweak corrections for these processes.


\section*{Note added:}
A user process for $gg \to HH$ based on \texttt{ggxy} has been implemented in the POWHEG-BOX framework, and can be obtained from
\begin{equation*}
\textrm{\url{https://gitlab.com/POWHEG-BOX/V2/User-Processes/ggxy_ggHH}.}
\end{equation*}


\section*{Acknowledgments}  

This research was supported by the Deutsche Forschungsgemeinschaft (DFG,
German Research Foundation) under grant 396021762 --- TRR 257 ``Particle
Physics Phenomenology after the Higgs Discovery''.
The work of K.~S.~was supported by the European Research Council (ERC)
under the European Union’s Horizon 2020 research and innovation programme
grant agreement 101019620 (ERC Advanced Grant TOPUP) and the UZH Postdoc Grant,
grant no.~[FK-24-115].
The work of J.~D.~was supported by STFC Consolidated Grant ST/X000699/1.
We thank Emanuele Bagnaschi for communications
concerning Ref.~\cite{Bagnaschi:2023rbx}
and Gudrun Heinrich for comments on the manuscript.

\begin{appendix}

\section{Low-level functions}
Beyond the ``high-level'' interface to the form factors described in
Section \ref{sec:ff-functions}, it is also possible to call the ``low-level''
functions which provide results for the exact (at one loop) and high-energy and small-$t$
expansions (at one and two loops) as well as the exact two-loop double-triangle contribution. In
these functions, the triangle contribution of $F_1$ is separated from the box contribution;
these pieces are called \texttt{FF0} and \texttt{FF1} in the function names, and $F_2$ is
called \texttt{FF2}.

The exact one-loop form factors can be evaluated using the functions defined in
the header file \texttt{ff/gghh/EXgghh1lFF.h},
\begin{lstlisting}
complex<double> EXgghh1lFF{0,1,2}(double s, double t,
                                  double mhs, double mts);
\end{lstlisting}
and the exact two-loop double-triangle contribution using the functions defined in
the header file \texttt{ff/gghh/DTgghh2lFF.h},
\begin{lstlisting}
complex<double> DTgghh2lFF{1,2}(double s, double t,
                                double mhs, double mts);
\end{lstlisting}
where although the triangle contribution is $t$-independent, each function
has the same signature. Here and below, the notation \texttt{EXgghh1lFF\{0,1,2\}}
implies that each of \texttt{EXgghh1lFF0}, \texttt{EXgghh1lFF1} and \texttt{EXgghh1lFF2}
is defined in the library.

At one and two loops, the high-energy and small-$t$ expansion results can be evaluated
using the functions defined in the header files
\texttt{ff/gghh/HEgghh$\{$1,2$\}$lFF.h} and \texttt{ff/gghh/t0gghh$\{$1,2$\}$lFF.h},
\begin{lstlisting}
complex<double> HEgghh{1,2}lFF{0,1,2}(double s, double t,
                                      double mhs, double mts);
\end{lstlisting}
where the Pad\'e approximation procedure has been already used, and
\begin{lstlisting}
complex<double> t0gghh{1,2}lFF{0,1,2}(double s, double t,
                                      double mhs, double mts);
\end{lstlisting}
which returns the sum of the small-$t$ expansion terms. At this level, the functions
should return numerically stable results within each expansion's region of validity,
but return nonsensical results beyond these regions.

At the ``lowest'' level, vectors of the expansion coefficients can be obtained.
At this level, no attempt is made to return sensible results in numerically unstable
regions or regions beyond the validity of the expansions. These functions are defined
in the header files \texttt{ff/gghh/HEgghh\{1,2\}lFF-lowlevel.h} and
\texttt{ff/gghh/t0gghh\{1,2\}lFF-lowlevel.h}.
For the high-energy expansion, the expansion coefficients for the real and imaginary
parts of each of the $m_H^2$ expansion terms are returned by
\begin{lstlisting}
vector<double> HEgghh{1,2}lFF{0,1,2}mhs{0,1,2}{re,im}(
                                        double s, double t,
                                        double mhs, double mts,
                                        unsigned mtsExpDepth = 24);
\end{lstlisting}
where the final parameter controls the depth at which the expansion is evaluated (at most 24).
The small-$t$ expansion coefficients are given by
\begin{lstlisting}
vector<complex<double>> t0gghh{1,2}lFF{0,1,2}mhs{0,1,2}(
                                        double s, double t,
                                        double mhs, double mts,
                                        unsigned tExpDepth = 6);
\end{lstlisting}
where again the final parameter controls the expansion depth (at most 6). At one loop,
higher-order $m_H^2$ terms are available: \texttt{mhs\{3,4\}}.

\end{appendix}



\bibliographystyle{jhep}
\bibliography{inspire.bib,extra.bib}

%


\end{document}